# Observation of room temperature magnetic skyrmions and their current-driven dynamics in ultrathin Co films

Seonghoon Woo,[1] Kai Litzius,[2,3] Benjamin Krüger,[2] Mi-Young Im,[4,5] Lucas Caretta,[1] Kornel Richter,[2] Maxwell Mann,[1] Andrea Krone,[2] Robert Reeve,[2] Markus Weigand,[6] Parnika Agrawal,[1] Peter Fischer,[4,7] Mathias Kläui,[2,3*] Geoffrey S. D. Beach[1*]

[1]Department of Materials Science and Engineering, Massachusetts Institute of Technology, Cambridge, Massachusetts 02139, USA

[2]Institut für Physik, Johannes Gutenberg-Universität Mainz, 55099 Mainz, Germany

[3]Graduate School of Excellence Materials Science in Mainz, Staudinger Weg 9, 55128 Mainz, Germany

[4]Center for X-ray Optics, Lawrence Berkeley National Laboratory, Berkeley, California, 94720, USA

[5]Daegu Gyeongbuk Institute of Science and Technology, Daegu 711-873, Korea

[6]Max Planck Institute for Intelligent Systems, 70569 Stuttgart, Germany

[7]Department of Physics, University of California, Santa Cruz, California 94056, USA

*Correspondence to:  gbeach@mit.edu, klaeui@uni-mainz.de

Magnetic skyrmions are topologically-protected spin textures that exhibit fascinating physical behaviors and large potential in highly energy efficient spintronic device applications. The main obstacles so far are that skyrmions have been observed in only a few exotic materials and at low temperatures, and manipulation of individual skyrmions has not yet been achieved. Here, we report the observation of stable magnetic skyrmions at room temperature in ultrathin transition metal ferromagnets with magnetic transmission soft x-ray microscopy. We demonstrate the ability to generate stable skyrmion lattices and drive trains of individual skyrmions by short current pulses along a magnetic racetrack. Our findings provide experimental evidence of recent predictions and open the door to room-temperature skyrmion spintronics in robust thin-film heterostructures.

Whereas the exchange interaction in common magnetic materials leads to collinear alignment of lattice spins, in materials with broken inversion symmetry and strong spin-orbit coupling, the Dzyaloshinskii-Moriya interaction (DMI) (*1*, *2*) can stabilize helical magnetic order (*3–9*). Of particular interest in such materials are magnetic skyrmions (*3*, *5*), particle-like chiral spin textures that are topologically protected from being continuously 'unwound.' Magnetic skyrmions can arrange spontaneously into lattices (*3*, *5*, *7–10*), and charge currents can displace them at remarkably low current densities (*11–17*). However, experimental observation of these intriguing and useful behaviors at room temperature has so far remained elusive.

Skyrmion lattices were first observed in B20 compounds such as MnSi (*5*, *7*), FeCoSi (*10*), and FeGe (*9*), whose non-centrosymmetric crystal structure gives rise to bulk DMI. Unfortunately, these materials order far below room temperature and only a few such compounds are known. DMI can also emerge at interfaces due to broken mirror symmetry (*18*). Spin cycloids and nanoscale skyrmion lattices have been observed in epitaxial ultrathin transition metal films by spin-polarized scanning tunnelling microscopy at ultralow temperatures (*6*, *8*). More recently, strong interfacial DMI in polycrystalline thin-film stacks such as Pt/CoFe/MgO, Pt/Co/Ni/Co/TaN, and Pt/Co/AlOx, has been inferred from field-induced asymmetries in domain wall (DW) motion (*19*, *20*) and nucleation (*21*). Micromagnetic simulations based on experimental estimates of DMI suggest that skyrmions can be nucleated, stabilized, and manipulated by charge currents in such materials (*17*), which could open the door to room-temperature skyrmion spintronics in robust thin-film heterostructures (*15*).

Here, we present direct observation of stable magnetic skyrmions and their current-driven motion in a thin transition metal ferromagnet at room temperature. Pt/Co/Ta multilayer stacks with perpendicular magnetic anisotropy were studied with high-resolution magnetic transmission soft x-ray microscopy (MTXM, Fig. 1A). Pt in contact with Co is known to generate strong DMI (*20*, *22*), while Ta generates very weak DMI (*22*–*24*), so that a large net DMI is anticipated in this asymmetric stack structure.

Figure 1B shows full-field MTXM images of the domain structure in a Pt/Co/Ta film, which allow the DMI strength to be quantified. At remanence the film exhibits a demagnetized labyrinth domain state, consistent with the sheared out-of-plane hysteresis loop in Fig. 1C, inset. With increasing out-of-plane field $B_z>0$, the up domains (dark contrast) grow while the down domains (light contrast) contract into narrow labyrinth domains, vanishing at saturation.

The domain width reflects the balance between the decreased demagnetizing energy and increased DW energy in the multidomain state. The latter scales with the DW surface energy density $\sigma_{DW} = 4\sqrt{AK_{u,eff}} - \pi|D|$, with $A$ the exchange stiffness, $K_{u,eff}$ the effective uniaxial anisotropy constant, and $D$ the DMI constant (*25*, *26*). Hence, $\sigma_{DW}$ is lowered and domain formation is more favorable whenever $D$ is finite. $\sigma_{DW}$ can be extracted from the domain widths using well-known domain spacing models (*27*), from which $|D|$ can be estimated.

The high spatial resolution (~25 nm) of MTXM allowed measurement of the domain widths $d_\uparrow$ and $d_\downarrow$ of up and down domains, respectively, as a function of increasing $B_z$ (Figure 1C) (*28*). At low fields, $d_\uparrow$ and $d_\downarrow$ vary linearly with $B_z$, maintaining a constant domain period $d = d_\downarrow + d_\uparrow \approx 480$ nm. At higher field $d_\uparrow$ diverges while $d_\downarrow$ approaches a terminal width $d_{\downarrow,min} \approx 100$ nm at $B_z \approx 10$ mT, showing little further variation until saturation. These behaviors are well-described by domain-spacing models for multilayer films (*27*), from which we find (*29*) $\sigma_{DW} = 1.3\pm0.2$ mJ/m². This is much smaller than $4\sqrt{AK_{u,eff}} \approx 5.3$ mJ/m² expected in the absence of DMI, implying a large $|D| = 1.3 \pm 0.2$ mJ/m$^2$, which is corroborated by micromagnetic simulations [see (*29*) for details].

Our micromagnetic simulations predict (*29*) the existence of a stable skyrmion lattice phase in this material on account of the strong DMI. We used scanning transmission x-ray microscopy (STXM) to confirm these predictions experimentally. Figure 2A shows STXM images of the domain structure in a 2 μm diameter Pt/Co/Ta disk during minor loop cycling of $B_z$. The left panel shows a parallel stripe phase at $B_z = -6$ mT, which transforms into a symmetric hexagonal skyrmion lattice after $B_z$ is swept to +2 mT, favoring up (dark-contrast) domains. This lattice closely resembles the micromagnetically-computed structure in (*29*), and the quasistatic transformation between the stripe phase and skyrmion lattice demonstrates that both these phases are metastable. Furthermore, our micromagnetic simulations predict that the skyrmion lattice is stable even at zero field for this material. To demonstrate this, we show in Figs. 2C,D the dynamical transformation of a labyrinth stripe domain phase into a hexagonal skyrmion lattice at $B_z = 0$, by applying short (6 ns) bipolar field pulses with a lithographically-patterned microcoil (Fig. 2B). The disk was first saturated in the up state, and then $B_z = -2$ mT was applied to nucleate a down labyrinth stripe domain (light contrast in Fig. 2C). The process of lattice formation can be seen as bipolar voltage pulse trains with increasing amplitude $V_{pp}$ are injected into the coil [see (*29*) for corresponding field profile]. As $V_{pp}$ is increased from 2 V to 9 V, the stripe domain begins to break into discrete skyrmions starting at one end (Fig. 2C) and after $V_{pp} = 10$ V, the domain structure has completely transformed into a geometrically confined skyrmion lattice. As the total domain area in the process does not change, the system is trying to find the most stable configuration while maintaining a fixed net magnetization.

The degree of order and the skyrmion size can be manipulated by low-amplitude pulse excitation (Fig 2D). After initializing the disk with a down labyrinth domain, $B_z$ was set to zero and a pulse train at $V_{pp}$ = 10 V was applied to create a stable array of skyrmions. By applying a small-amplitude pulse train $V_{pp}$ = 4 V, the skyrmions relax into a highly-ordered hexagonal lattice, without changing their size. Increasing slightly the pulsed field amplitude to $V_{pp}$ = 5 V (Fig. 2D) decreases the skyrmion size and increases their density, showing that multiple skyrmion lattice periodicities can be stabilized if commensurate with the confining geometry.

Having established that skyrmion lattices can be stabilized in this material, we next investigate their manipulation by current in a magnetic track. Recent simulations (*17*) suggest that skyrmions in ultrathin films might be driven even more efficiently than in previous studies by vertical spin current injection, which can occur when charge current flows in an adjacent heavy metal due to the spin Hall effect. In this case the current exerts a Slonczewski-like torque that can be much stronger than spin-transfer torques from in-plane spin current. When Pt and Ta with large spin Hall angles are placed on opposite sides of a ferromagnet, the spin Hall currents generated at each interface work in concert to generate a large Slonczewski-like torque (*30*). As the spin Hall effect-direction of motion of a skyrmion depends on its topology (*17*), observations of current-induced displacement can serve to verify the topology and chirality of the skyrmions in this system.

An external magnetic field $B_z$ = -22 mT was applied to a 2 μm-wide, 300 μm-long track contacted by Au electrodes at either end for current injection (Fig. 3A). $B_z$ causes the initial labyrinth domains to shrink into a few isolated skyrmions. Figure 3B shows a sequence of STXM images of a train of four skyrmions stabilized by $B_z$. Each image was acquired after injecting 20 current pulses with current-density amplitude $j_a = 2.2 \times 10^{11}$ A/m² and duration 20 ns, with the polarity indicated in the figure. Three of the four skyrmions move freely along the track, and can be displaced forward and backward by current, while the left-most skyrmion remains immobile, evidently pinned by a defect. The propagation direction is along the current flow direction (against electron flow), and this same directionality was observed for skyrmions of opposite polarity. Micromagnetic simulations (*29*) show that the observed unidirectional spin-Hall driven displacement is consistent with Néel skyrmions with left-handed chirality, confirming the topological nature of the skyrmions in this material.

We found that the four skyrmions do not all move at the same speed, suggesting a significant influence of pinning on the skyrmion motion in this current regime. At the highest current density used, we find that pinned skyrmions can be annihilated, as in the last image of Fig. 3B, where only three skyrmions remain, and the leftmost skyrmion becomes pinned at the same location as was the annihilated skyrmion. The average velocity of the most mobile skyrmion versus current density is shown in Fig 3C, with the narrow range determined by the critical propagation threshold and the maximum pulse amplitude experimentally available. We find velocities that are significantly lower than calculated for a defect-free sample (*29*), in contrast to recent micromagnetic studies that predict high skyrmion mobility even in the presence of discrete defects (*14–17*). Our micromagnetic simulations (*29*) suggest that dispersion in local magnetic material parameters at the nanometer lenghtscale can cause skyrmion pinning and leads to reduced velocities qualitatively consistent with experiments.

Our results show that in common polycrystalline transition metal ferromagnets magnetic skyrmions and skyrmion lattices can be stabilized and manipulated in confined geometries at room temperature. Since the magnetic properties of thin-film heterostructures can be tuned over a wide range by varying layer thicknesses, composition, and interface materials, this work highlights the possibility to engineer the properties of skyrmions and their dynamics using materials that can be readily integrated into spintronic devices.

**Methods:**

The Pt/Co/Ta films were grown by dc magnetron sputtering at room temperature under 3 mTorr Ar at a background pressure of ~$2\times10^{-7}$ Torr. A thin Ta seed layer was deposited prior to growing the multilayer stack. 15 layer repetitions were grown to enhance magnetic contrast in the x-ray magnetic circular dichroism (XMCD) signal and to more readily destabilize the uniformly-magnetized state on account of the increased demagnetizing field. Samples were grown on 100 nm-thick $Si_3N_4$ membranes for XMCD imaging in transmission geometry. Nominally-identical companion films were grown on thermally-oxidized Si wafers for magnetic characterization by vibrating sample magnetometry (VSM). VSM measurements yield an in-plane saturation field $\mu_0 H_k = 0.5$ T, and a saturation magnetization $M_s = 6\times10^5$ A/m, normalized to the nominal Co volume.

The disk and track structures in Figure 2 and Figure 3 were patterned using electron beam lithography and lift-off. The microcoil in Figure 2 and current injection contacts in Fig. 3 were sputtered Ti(5nm)/Au(100nm) bilayers patterned using a second lithography step.

Samples for XMCD imaging were oriented with the surface normal parallel to the circularly-polarized x-ray beam (main text Fig. 1A), so that the XMCD contrast is sensitive to the out-of-plane magnetization component. The images in Figure 1 of the main text were acquired using full-field MTXM performed at the XM-1 beamline 6.1.2 at the Advanced Light Source in Berkeley, California. The images in Figures 2 and 3 of the main text were acquired using scanning transmission x-ray microscopy (STXM) at the MAXYMUS beamline at the BESSY II synchrotron in Berlin.

The bipolar field pulses applied using the microcoil in Figure 3 of the main text were generated by injecting square bipolar voltage pulses with peak-to-peak amplitude $V_{pp}$, and a length of 6 ns. The pulse trains referred to in the main text consisted of 30 s applications of these bipolar pulses at a repetition rate of 1.43 MHz. A peak-to-peak voltage amplitude $V_{pp} = 1V$ corresponds to a maximum current amplitude of 6 mA, or a current density of $6\times10^{10}$ A/m$^2$. Details of the coil geometry and calculated magnetic field profile are discussed in the Supplementary.

**References:**
1. I. Dzyaloshinsky, *J. Phys. Chem. Solids*. **4**, 241–255 (1958).
2. T. Moriya, *Phys. Rev.*. **120**, 91–98 (1960).
3. U. K. Rößler, A. N. Bogdanov, C. Pfleiderer, *Nature*. **442**, 797–801 (2006).
4. M. Uchida, Y. Onose, Y. Matsui, Y. Tokura, *Science*. **311**, 359–361 (2006).
5. S. Mühlbauer *et al.*, *Science*. **323**, 915–919 (2009).
6. M. Bode *et al.*, *Nature*. **447**, 190–193 (2007).
7. X. Z. Yu *et al.*, *Nature*. **465**, 901–904 (2010).
8. S. Heinze *et al.*, *Nat. Phys.*. **7**, 713–718 (2011).
9. X. Z. Yu *et al.*, *Nat. Mater.*. **10**, 106–109 (2011).
10. W. Münzer *et al.*, *Phys. Rev. B*. **81**, 041203 (2010).
11. F. Jonietz *et al.*, *Science*. **330**, 1648–1651 (2010).
12. T. Schulz *et al.*, *Nat. Phys.*. **8**, 301–304 (2012).
13. X. Z. Yu *et al.*, *Nat. Commun.*. **3**, 988 (2012).
14. J. Iwasaki, M. Mochizuki, N. Nagaosa, *Nat. Commun.*. **4**, 1463 (2013).
15. A. Fert, V. Cros, J. Sampaio, *Nat. Nanotechnol.*. **8**, 152–156 (2013).

16. J. Iwasaki, M. Mochizuki, N. Nagaosa, *Nat. Nanotechnol.*. **8**, 742–747 (2013).

17. J. Sampaio, V. Cros, S. Rohart, A. Thiaville, A. Fert, *Nat. Nanotechnol.*. **8**, 839–844 (2013).

18. A. Fert, P. M. Levy, *Phys. Rev. Lett.*. **44**, 1538–1541 (1980).

19. S. Emori, U. Bauer, S.-M. Ahn, E. Martinez, G. S. D. Beach, *Nat. Mater.*. **12**, 611–616 (2013).

20. K.-S. Ryu, L. Thomas, S.-H. Yang, S. Parkin, *Nat. Nanotechnol.*. **8**, 527–533 (2013).

21. S. Pizzini *et al.*, *Phys. Rev. Lett.*. **113**, 047203 (2014).

22. S. Emori *et al.*, *Phys. Rev. B*. **90**, 184427 (2014).

23. J. Torrejon *et al.*, *Nat. Commun.*. **5** (2014), doi:10.1038/ncomms5655.

24. R. Lo Conte *et al.*, *Phys. Rev. B*. **91**, 014433 (2015).

25. M. Heide, G. Bihlmayer, S. Blügel, *Phys. Rev. B*. **78**, 140403 (2008).

26. A. Thiaville, S. Rohart, É. Jué, V. Cros, A. Fert, *EPL Europhys. Lett.*. **100**, 57002 (2012).

27. Z. Malek, V. Kambersky, *Czeck J Phys*. **8**, 416 (1958).

28. The large apparent remanent magnetization and coercivity (field where up and down domains are the same width) suggested by the low-field MTXM images likely arise from a large remanent field in the electromagnet, which has not been corrected for.

29. See supplementary materials

30. S. Woo, M. Mann, A. J. Tan, L. Caretta, G. S. D. Beach, *Appl. Phys. Lett.*. **105**, 212404 (2014).

**Acknowledgments:**
This material is based upon work supported by the U.S. Department of Energy, Office of Science, Office of Basic Energy Sciences under Award Number DE-SC0012371. G.B. acknowledges support from C-SPIN, one of the six SRC STARnet Centers, sponsored by MARCO and DARPA. P.F. and M-Y.I acknowledge support from the Director, Office of Science, Office of Basic Energy Sciences, Materials Sciences and Engineering Division, of the U.S. Department of Energy under Contract No. DE-AC02-05-CH11231 and the Leading Foreign Research Institute Recruitment Program (Grant No. 2012K1A4A3053565) through the National Research Foundation of Korea (NRF) funded by the Ministry of Education, Science and Technology (MEST). M.K. acknowledges support by the DFG, the Graduate School of Excellence Materials Science in Mainz (MAINZ) GSC 266 the EU (MASPIC, ERC-2007-StG 208162; WALL, FP7-PEOPLE-2013-ITN 608031; MAGWIRE, FP7-ICT-2009-5), and the Research Center of Innovative and Emerging Materials at Johannes Gutenberg University (CINEMA). BK is grateful for financial support by the Carl-Zeiss-Foundation. KL gratefully acknowledges financial support by the Graduate School of Excellence Materials Science in Mainz (MAINZ). Measurements were carried out at the MAXYMUS endstation at Helmholtz-Zentrum Berlin. We thank HZB for the allocation of synchrotron radiation beamtime.

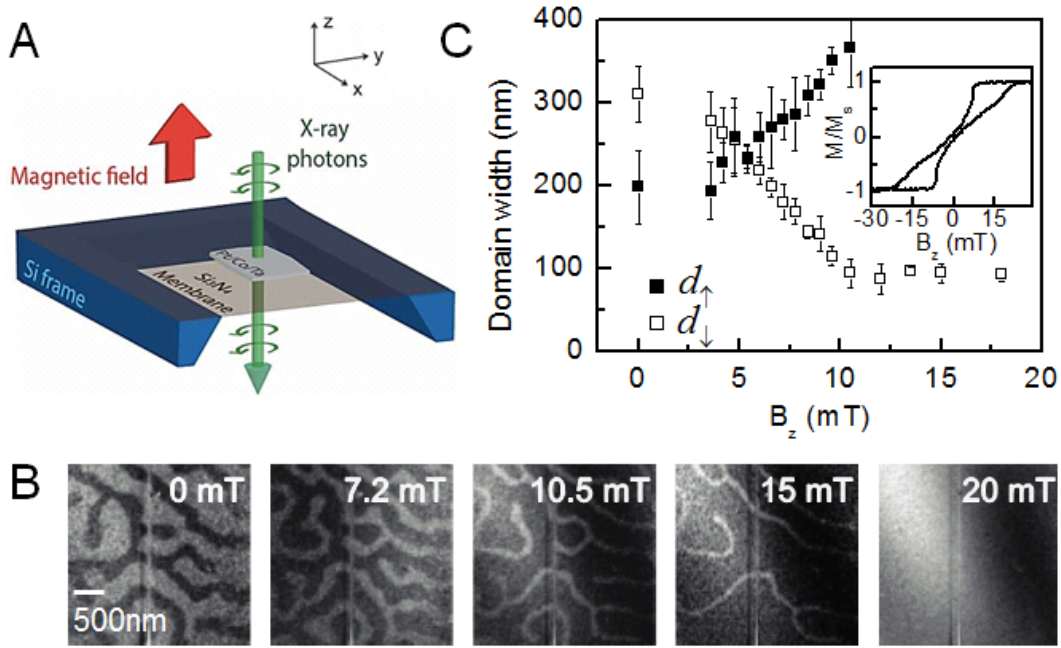

**Fig. 1**. **Soft X-ray imaging of domain structure.** (**A**) Schematic of magnetic transmission soft x-ray microscopy (MTXM) geometry. (**B**) Series of MTXM images acquired after negative field saturation for several increasing fields $B_z>0$ for a continuous Pt/Co/Ta film. Dark and light contrast correspond to magnetization oriented up (along +z) and down (along -z), respectively. (**C**) Width of up and down domains versus $B_z$, determined from MTXM images. Error bars denote standard deviation of ten individual width measurements. Inset in (**C**) shows a hysteresis loop for a companion film grown on a Si wafer.

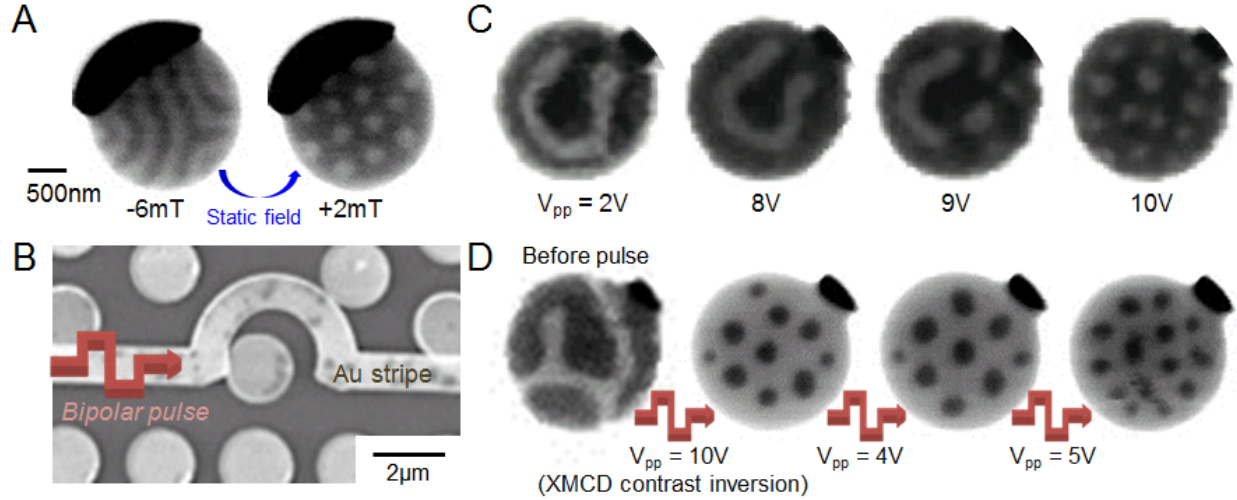

**Fig. 2. Skyrmion lattice generation. (A)** STXM images of the domain state in a 2 μm Pt/Co/Ta disk at $B_z$ = -6 mT (left) and after subsequently applying $B_z$ = 2 mT (right). **(B)** Scanning electron micrograph of magnetic disk array with an Au microcoil patterned around one disk. **(C)** Sequence of STXM images after applying bipolar pulse trains (peak-to-peak voltage amplitude $V_{pp}$) with the microcoil, showing transformation from labyrinth stripe domain into skyrmion lattice. **(D)** An initial labyrinth domain state was generated by static field (first image) and then transformed into a hexagonal skyrmion lattice by applying a bipolar pulse train with $V_{pp}$ = 10 V (second image). The last two images were acquired after applying $V_{pp}$ = 4 V and $V_{pp}$ = 5 V, respectively. Dark (light) contrast corresponds to up (down) magnetization in all STXM images except for the last three in **(D)**, where the XMCD contrast was inverted.

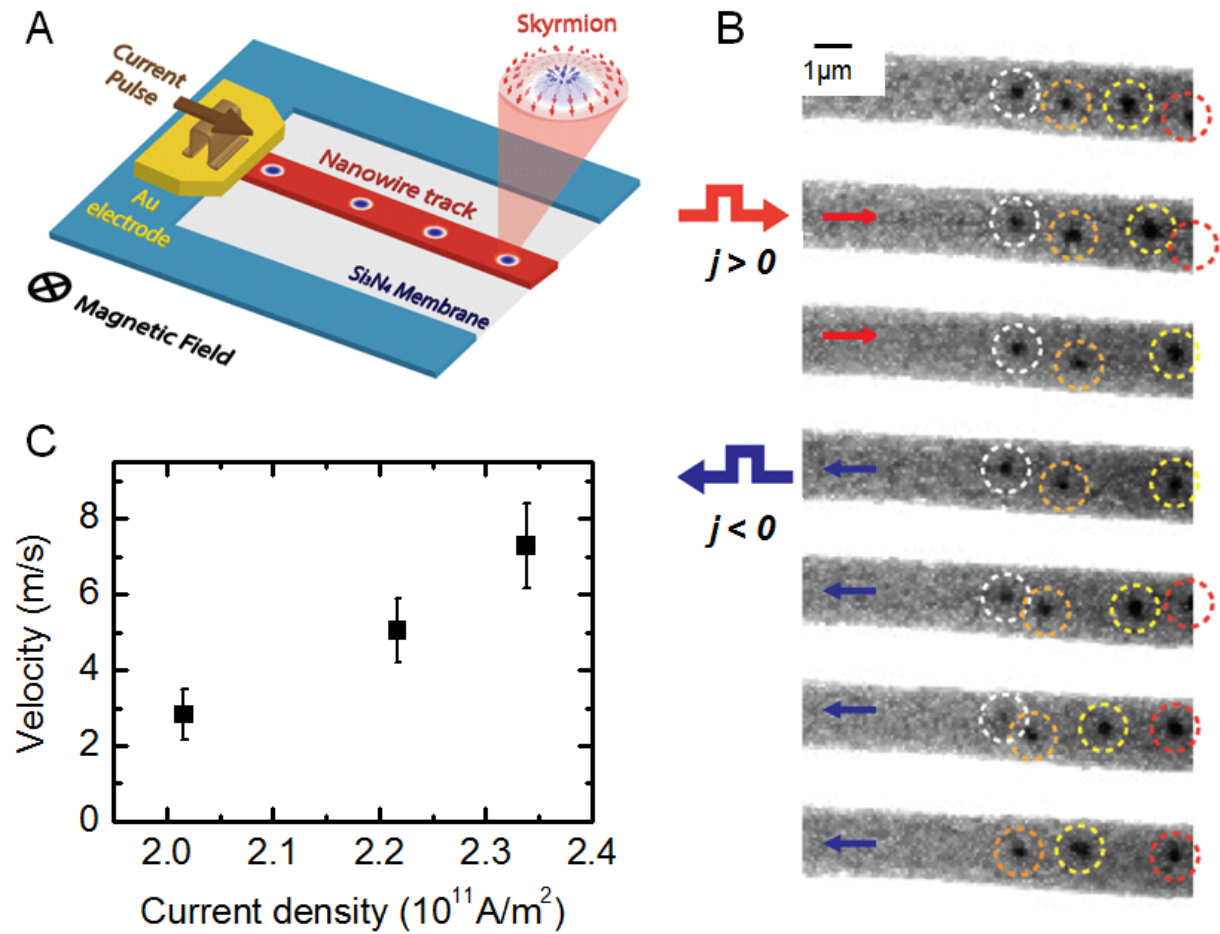

**Fig. 3. Current-driven skyrmion motion.** **(A)** Schematic of magnetic track on $Si_3N_4$ membrane with current contacts and skyrmions stabilized by a down-directed applied magnetic field. **(B)** Sequential STXM images showing skyrmion displacement after injecting 20 unipolar current pulses along the track, with an amplitude $j_a = 2.2 \times 10^{11}$ and polarity as indicated. Individual skyrmions are outlined in colored circles for clarity. **(C)** Average velocity versus current density for the skyrmion circled in yellow in **(B)**.